\documentclass[amsmath,amssymb,amssymb,aps,pra, showpacs,preprint,longbibliography]{revtex4-1}

\usepackage{graphicx}% Include figure files
\usepackage{mathrsfs}
\usepackage{dcolumn}% Align table columns on decimal point
\usepackage{bm}% bold math
\usepackage{natbib}

\newcommand{\bra}[1]{\ensuremath{\left\langle #1\right|}}
\newcommand{\ket}[1]{\ensuremath{\left|#1\right\rangle}}

\begin{document}
\title{Time-, Frequency-, and Wavevector-Resolved X-Ray Diffraction from Single Molecules}
\author{Kochise Bennett}
\email{Email:kcbennet@uci.edu}
\author{Jason D. Biggs}
\author{Yu Zhang}
\author{Konstantin E. Dorfman}
\author{Shaul Mukamel}
\email{smukamel@uci.edu}
\affiliation{University of California, Irvine, California 92697-2025}
\date{\today}

\begin{abstract}
Using a quantum electrodynamic framework, we calculate the off-resonant scattering of a broad-band X-ray pulse
from a sample initially prepared in an arbitrary superposition of electronic states.  The signal consists of
single-particle (incoherent) and two-particle (coherent) contributions that carry different particle form
factors that involve different material transitions.  Single-molecule experiments involving incoherent
scattering are more influenced by inelastic processes compared to bulk measurements. The conditions under
which the technique directly measures charge densities (and can be considered as diffraction) as opposed to correlation functions of the charge-density are
specified. The results are illustrated with time- and
wavevector-resolved signals from a single amino acid molecule (cysteine) following an impulsive excitation by
a stimulated X-ray Raman process resonant with the sulfur K-edge.  Our theory and simulations can guide future
experimental studies on the structures of nano-particles and proteins.
\end{abstract}

\maketitle

\section{Introduction}

X-ray techniques have long been applied to image the electronic charge density of atoms, molecules, and
materials \cite{bressler2004ultrafast, velser, rixsament}.  Recently-developed X-ray free electron laser
sources, which generate short (attosecond), intense pulses, open up numerous potential applications for high
temporal and spatial resolution studies \cite{naturefsnano, altarelli, feldhaus, mcneil, siders, Miller07032014}.  One
exciting application is the determination of molecular structure by X-ray diffraction of nanocrystals
\cite{naturefsnano, koopvivo} avoiding the crystal growth process which is often the bottleneck in structure
determination \cite{stevens, mcpherson}; it may take decades to crystallize a complex protein.  It is much
easier to grow nanocrystals than the many-microns-sized samples required by conventional crystallography. This
has been demonstrated experimentally in nanocrystals for the water splitting photosynthetic complex II
\cite{kernroom}, a mimivirus \cite{seibert}, and a membrane protein \cite{naturefsnano, fromme}.
\par
Extending this idea all the way to the single-molecule level, totally removing ``crystal" from
crystallogrpaphy is an intruiging possibility \cite{hajdu, chapmanbeyond, starodub}.  Obtaining a protein
structure by scattering from a single molecule is revolutionary.  Many obstacles need to be overcome to
accomplish this ambitious goal.  For instance, the molecule will typically break down when subjected to such
high fluxes.  However, it has been argued \cite{hajdu, chapmanfs, schlichting, neutze} that, for sufficiently
short pulses, the scattering occurs prior to photon damage, so that this should not affect the measured charge
density. This point is still under debate.
\par
In this paper, we consider the scattering of a broad-band X-ray pulse from a system composed of a single, few,
or many molecules prepared in a superposition of electronic states and show under what conditions the signal
may be described solely by the time-dependent charge density.  We assume that the X-ray pulse is off-resonance
from any material transitions and that the pulse is sufficiently short so that the electrons in the sample do
not appreciably re-arrange during the pulse time.  The fluence is assumed sufficiently low so that the
scattering is linear in pulse intensity.  Scattering off non-stationary evolving states is an area of growing
interest. When the time-dependent state of matter merely follows some classical parameter (as in the case of
tracking the time-dependent melting of crystals \cite{siders, falcone}) no coherence of electronic states is
prepared and the analysis of scattering is simplified considerably. Time-dependent diffraction can then be
described by simply replacing the charge density in stationary diffraction by the time-dependent
charge-density.  The situation is more complex in pump-probe experiments in which a superposition of
electronic states is prepared by the pump and is then probed by X-ray scattering. Such superpositions, which
involve electronic coherence, can be prepared e.g. by inelastic stimulated Raman processes
\cite{annualReview}, a photoionization process \cite{cederbaum}, off-resonant femtosecond pulses
\cite{bargheer} or high-intensity optical pumping \cite{stingl}. Diffraction is a macroscopic, classical
effect that involves the interference of wavefronts emanating from different sources treated at the level of
Maxwell's equations \cite{modxrayphys}. A fundamental difficulty in extending it to single molecules is that
X-ray scattering (as any light scattering) from a single molecule may not be thought of simply as a
diffraction since it has both elastic (Thompson/Rayleigh) and an inelastic (Compton/Raman) components
\cite{dorfman}. Using a quantum electrodynamic (QED) approach, we discuss and analyze the single-particle vs.
the cooperative contributions to the signal.  We find that the two terms carry different particle form factors
that permit Raman scattering in different frequency ranges.  Simulations are presented for the scattering of a
broad-band X-ray pulse from a single molecule of the amino acid cysteine either in the ground state or when
prepared in a nonstationary state by a stimuated resonant X-ray Raman process with various delay times between
preparation and the scattering event.

\section{Classical Theory of Diffraction}
As the expressions developed in this paper bear a resemblence to the standard classical theory of diffraction,
we review it briefly.  The diffraction signal from a system initially in the ground state $\vert g\rangle$ is
\begin{equation}\label{eq:SkClassic}
S(\mathbf{q})\propto\vert\sigma_{gg}(\mathbf{q})\vert^2,
\end{equation}
where $\sigma_{gg}(\mathbf{q})=\langle g\vert \hat{\sigma}(\mathbf{q})\vert g\rangle$ is the ground state charge
density in $\mathbf{q}$-space and $\mathbf{q}\equiv \mathbf{k}_s-\mathbf{k}_p$ is the momentum transfer ($\mathbf{k}_s$
is the outgoing mode and $\mathbf{k}_p$ is the incoming mode).  More generally, $\hat{\sigma}(\mathbf{q})$ is the Fourier
transform of the charge-density operator
\begin{equation}\notag
\hat{\sigma}(\mathbf{q})=\int d\mathbf{r}\hat{\sigma}(\mathbf{r})e^{-i\mathbf{q}\cdot\mathbf{r}}.
\end{equation}
Equation (\ref{eq:SkClassic}) assumes that the scattering is elastic and treats the entire sample as a single system
(i.e. $\hat{\sigma(\mathbf{r})}$ is the electron density of the entire sample).  A common approach to understanding
diffraction patterns involves making an approximation on the structure of the electron density.  For example, we
may presume that the complete wavefunction is built up from single-electron wavefunctions and the total
electron density is the sum of the densities associated with each electon wavefunction.  If the system
is composed of $N$ identical particles and each electron is bound to a particle, then the amplitude of
the signal from each particle carries a relative phase related to the particle position.  The term
``particle" is used here as a generic partitioning of the system and may stand for atoms as well as
large (molecules) or small (unit cell) groups of atoms.  Practically speaking, they should be large
enough that the electron density between particles may be safely neglected and small enough that
electronic structure calculations can be performed.  Separating single- and multi-particle contributions
yields for the intensity \cite{modxrayphys}:
\begin{equation}\label{eq:SkClassicSep}
S(\mathbf{q})\propto\vert\sigma_{aa}(\mathbf{q})\vert^2[N+\sum_{\alpha\ne \beta}e^{-i\mathbf{q}\cdot\mathbf{r}_{\alpha\beta}}]
\end{equation}
where $\mathbf{r}_{\alpha\beta}\equiv\mathbf{r}_\alpha-\mathbf{r}_\beta$ is the position of
particle $\alpha$ relative to $\beta$. Here, $\sigma_{aa}(\mathbf{r})$ stands for the electron
density in a single particle and $\sigma_{aa}(\mathbf{q})$ is known as the particle form factor.
Thus, the signal is factored into a product of the signals from the particle electron density and
from the distribution of particles.
The two terms in equation (\ref{eq:SkClassicSep}) both contribute linearly in $N$ to the integrated
signal \cite{guiner}.  However, the former yields a signal that is generally distributed throughout
reciprocal space while the various terms in the $\alpha\ne \beta$ summation carry different phases
that cause a redistribution of the signal into points of constructive and destructive interference
(a Bragg pattern for crystaline samples) \footnote{Spots of constructive interference (known as the speckle pattern)
are determined by the interparticle structure factor \unexpanded{$\sum_{\alpha\ne \beta}e^{-i\mathbf{q}_{aa}\cdot\mathbf{r}_{\alpha\beta}}$}.
The peak intensity at the points of constructive interference (such as the Bragg peaks for crystals or small angle scattering
(SAXS)) scales as \unexpanded{$N^2$}.
This scaling facilitates the measurement of the scattered intensity at Bragg peaks or for small
\unexpanded{$\mathbf{q}$}.\cite{miao2004}}.

\section{Off-Resonance X-ray Scattering Signals}\label{sec:loops}
\begin{figure}
\includegraphics[width =0.75\textwidth]{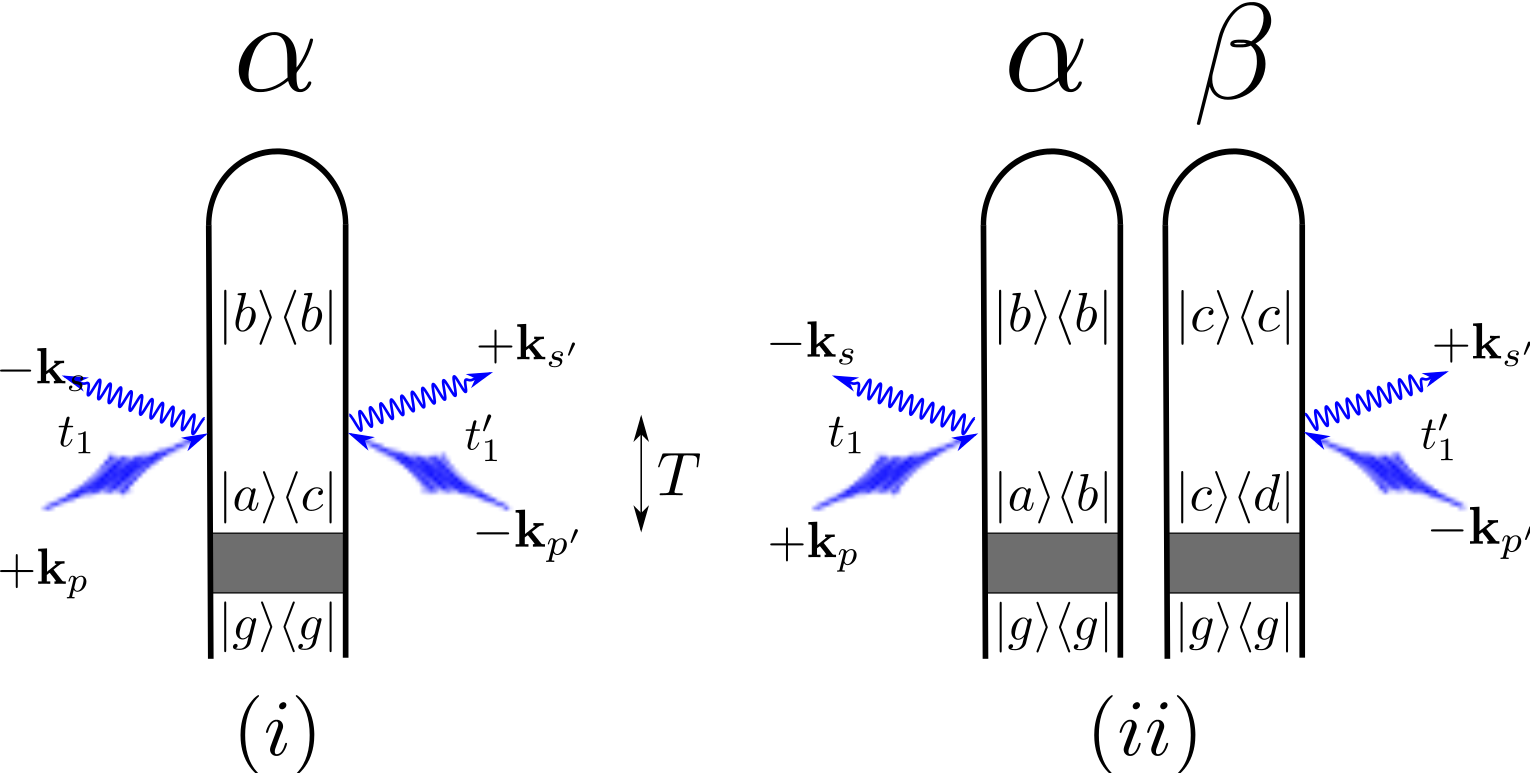}
\caption{Loop diagrams for incoherent (\textbf{a}) and coherent (\textbf{b}) X-ray scattering processes.  The shaded
area represents an unspecified process that prepares the system in an arbitrary state ($\vert g\rangle$ is the
electronic ground state).  We denote modes of the pump with $p$ and $p'$ whereas $s$, $s'$ represent relevant
scattering modes ($\mathbf{k}_{p^{(\prime)}}$ has frequency $\omega_{p^{(\prime)}}$ and
$\mathbf{k}_{s^{(\prime)}}$ has frequency $\omega_{s^{(\prime)}}$). The time $T$ between the termination of
this preparation process and the central time of the scattered pulse is shown via the arrow in the center of
the figure.  Elastic scattering corresponds to $\omega_{ab}=\omega_{bc}=0$ (i.e. $\omega_{ac}=0$ for the
incoherent and $\omega_{bc}=\omega_{ed}=0$ for the coherent contribution.  Elastic scattering therefore
originates from scattering off populations. For diagram rules, see \cite{dorfman, biggs_coherent_2011}}
\label{fig:1D}
\end{figure}

We calculate the X-ray scattering by a sample initially prepared in a non-stationary state using a quantum
description of the field, in which radiation back-reaction (which leads to inelastic scattering) is naturally
built in. Incorporation of the broad frequency bandwidth of incoming X-rays is an important point as
ultra-short pulses need to outrun destruction in single-molecule scattering.  We assume that the molecule is
initially prepared in an arbitrary density matrix, representing a pure or mixed state such as photo-ions. We
extend the quantum field formalism developed in \cite{tanaka, dorfman} for spontaneous emission of visible
light to off-resonant X-ray scattering.  Our calculation starts with the minimal-coupling Hamiltonian, which
contains a term proportional to $\hat{\mathbf{A}}^2\hat{\sigma}$ as well as
$\hat{\mathbf{j}}\cdot\hat{\mathbf{A}}$ where $\hat{\mathbf{j}}$ is the electronic current operator. The former
describes instantaneous scattering where the electrons do not have the time to respond during the scattering
process.  The second term dominates resonant processes and allows for a delay between absorption and emission
during which electronic rearrangement, ionization, and breakdown can occur.  Since hard
X-rays are always resonant with the electronic continua representing various ionized states, the relative role
of the $\hat{\mathbf{j}}\cdot\hat{\mathbf{A}}$ term should be investigated further in order to clarify how
important are these processes.  In our study, we treat only off-resonant scattering and therefore focus on the
$\hat{\mathbf{A}}^2$ term, which dominates such processes \cite{karle1980some, schulke}.
\par
In the interaction picture, the Hamiltonian is given by
\begin{equation}
\hat{H}=\hat{H}_0+\hat{H}'(t),
\end{equation}
where $\hat{H}_0$ is the bare field and matter Hamiltonian while $\hat{H}'(t)$ is the field-matter coupling.  Assuming the diffracting X-ray pulse is not resonant with any material transitions and its intensity is not too high, its interaction with the matter is given by \cite{schulke}
\begin{equation}\label{eq:Hprime}
\hat{H}'(t)=\frac{1}{2}\int d\mathbf{r}\hat{\mathbf{A}}^2(\mathbf{r},t)\hat{\sigma}_T(\mathbf{r},t).
\end{equation}
We consider a sample of $N$ non-interacting, identical particles (molecules or atoms) indexed by $\alpha$ with non-overlapping charge distributions so that the total charge-density operator can be partitioned as $\hat{\sigma}_T(\mathbf{r})=\sum_{\alpha}\hat{\sigma}(\mathbf{r}-\mathbf{r}_\alpha)$. The signal is given by the electric field intensity arriving at the detector and may be generally expressed as the overlap integral of a detector spectrogram (given in terms of the detection parameters) and a bare spectrogram (equation \ref{eq:Sspectrograms}).
\par
The scattering signal mode is initially in the vacuum state $\vert 0\rangle\langle 0\vert$. Therefore an interaction on both bra and ket of the field density matrix is required to generate the state $\vert 1\rangle\langle 1\vert$ which gives the signal. We thus expand the signal as
\begin{equation}\label{eq:Sgatedintensity}
S(\bar{\omega},\bar{t},\bar{\mathbf{r}},\bar{\mathbf{k}})=\int dt\int d\mathbf{r}\langle\mathbf{E}^{(t\mathbf{r}f\mathbf{k})\dagger}(\mathbf{r},t)\mathbf{E}^{(t\mathbf{r}f\mathbf{k})}(\mathbf{r},t)\rangle.
\end{equation}
It is given in terms of the gated electric fields (defined in Appendix \ref{ap:gating}), to second order in $\hat{H}'$ (equation \ref{eq:Hprime}). This naturally leads to a double sum $\sum_{\alpha,\beta}$ over the scatterers. Terms with $\alpha=\beta$ arise when the probe pulse is scattered off a single particle (Fig.~\ref{fig:1D}.\textbf{a}) and terms with $\alpha\ne\beta$ describe two-particle scattering events (Fig.~\ref{fig:1D}.\textbf{b}). The former contains $N$ terms which add incoherently (at the intensity level), giving a virtually isotropic signal. The latter, in contrast, is governed by $N(N-1)$ terms which carry different phase-factors and can interfere destructively or constructively (yielding the Bragg peaks in a crystalline sample or, more generally, a speckle pattern) \cite{modxrayphys}. In general, both contributions must be considered as it is frequently not sufficient to sample a signal only at the points of constructive interference (Bragg diffraction) where the two-particle terms dominate \cite{miao2004}. In this paper, we refer to the single-particle contribution as ``incoherent'' and the two-particle contribution as ``coherent" reflecting the way in which these contributions add (at the intensity versus amplitude level). In the X-ray community, incoherent is commonly taken to refer to inelastic contributions while coherent refers to elastic. The two nomenclatures coincide when considering scattering from the ground state because, as will be shown below, two-particle scattering from the ground state (or any population) is necessarily elastic while single-particle scattering from the ground state has only a single elastic term (or one term for each initially populated state). Two-particle scattering from a nonstationary superposition state, in contrast, has both elastic and inelastic terms which add coherently (since they are both proportional to a spatial phase-factor $e^{i\Delta\mathbf{k}\cdot\mathbf{r}_{ab}}$ dependent on the distance between the two particles). On the other hand, the elastic terms from the single-particle scattering add incoherently since their spatial phase factors are canceled by the opposite-hermiticity interactions on the ket/bra.
\par
In the coherent terms the product of charge densities on different particles can be factorized and the signal is proportional to the modulus square of the single-particle electron density $\hat{\sigma}(\mathbf{r})$. Such factorization is not possible for the incoherent terms where the signal is given by a correlation function of the charge density rather than the charge density itself.  As discussed below, restricting attention to elastic scattering reduces the correlation function to the modulus-square result and eliminates the need to consider the correlation function.  Thus, the signal is expressible in terms of the charge-density alone either when the coherent contribution dominates or when attention is restricted to elastic scattering.
\par
Figure \ref{fig:1D} illustrates the incoherent (Fig.~\ref{fig:1D}.\textbf{a}) and coherent (Fig.~\ref{fig:1D}.\textbf{b})  scattering processes from a sample following preparation in a non-stationary electronic superposition state described by the density matrix $\hat{\rho}=\sum_{cb}\rho_{cb}\vert c\rangle\langle b\vert$.  The preparation process is represented by the gray box.  In this paper, we will consider a Rama preparation process (in another work, we examine the case where the preparation process is itself an off-resonant scattering process \cite{jpbpaper}).  After the preparation process (which terminates at $t=0$) the system evolves freely until an X-ray probe pulse with an experimentally-controlled envelope centered at time $T$ impinges on the sample and is scattered into a signal mode that is initially in a vacuum state.  This signal photon is then finally absorbed by the detector.  Time translation invariance of the matter correlation function implies the basic energy-conservation condition $\omega_p-\omega_{p'}+\omega_{s'}-\omega_{s}=\omega_{ac}$ for an incoherent scattering event (Fig.~\ref{fig:1D}.\textbf{a}). Coherent scattering events (Fig.~\ref{fig:1D}.\textbf{b}) give two such conditions corresponding to the two diagrams $\omega_p-\omega_s=\omega_{ab}$ and $\omega_{p'}-\omega_{s'}=\omega_{cd}$.
For a broad-band pulse in which $\omega_p$ and $\omega_{p'}$ can differ appreciably, the signal will contain contributions from paths in which $\omega_s$ and $\omega_{s'}$ take all possible values within this bandwidth.  These can be controlled by pulse-shaping techniques as well as by the choice of detection parameters \cite{dorfman1}.
\par
It follows from the diagrams that, since each particle must end the process in a population state and has only interactions on one side of the loop, coherent scattering from populations will always be Rayleigh type while Raman (inelastic) processes result from scattering off coherences.  In contrast, the single-particle energy conservation condition for an initial population (i.e. $\omega_{ac}=0$) allows for inelastic processes.

\section{ Frequency-Gated Signals}\label{sec:signals}
Complete expressions for the signal which include arbitrary gating and pulse envelopes (i.e., the bare spectrograms) are given in the Appendix \ref{ap:barederivation}. In this section, we discuss the simpler signal that results when no time-gating is applied.  The formulas are simplest when we employ delta functions for the detector spectrograms.  As shown Appendix \ref{ap:gating} we have separate detector spectrograms for the time-frequency gating and the space-propagation gating.  As seen in Appendix \ref{ap:barederivation} (equations (\ref{eq:WBcohCF}) and (\ref{eq:WBincCF})) both coherent and incoherent bare spectrograms carry the delta function factor $\delta(\mathbf{k}'-\frac{\omega'}{c}\hat{\mathbf{r}}')$.  This connects $\omega'$ to $\mathbf{k}'$ in the usual way (though this is not automatic since the two are not a priori related in this way but rather both begin as seperate gating variables) as well as fixing the direction of $\mathbf{k}'$.  For this reason, the logical choice for the spatial-propagation detector spectrogram is
\begin{equation}
W_D(\mathbf{r}',\mathbf{k}';\bar{\mathbf{r}},\bar{\mathbf{k}})=\delta(\mathbf{r}'-\bar{\mathbf{r}})
\end{equation}
This represents a spatially resolved signal; that is, the location of the detection event (the pixel location) is resolved.  All signals considered in this paper use this choice.
\par
If the detector spectrogram does not depend on $t'$ (no time-gating is applied), we may separate the time-dependent phase factors from the auxilliary functions and carry out the time integration to give a factor of $\delta(\omega_p-\omega_{p'}+\tilde{\omega}'-\tilde{\omega})$.  The signals are therefore given by
\begin{widetext}
\begin{align}\label{eq:Scohwr}
S_{\mathrm{coh}}(\bar{\omega},\bar{\mathbf{r}},\Lambda)=K\int d\omega'\vert F_f(\omega',\bar{\omega})\vert^2 \omega'^2\sum_{\alpha\beta}\int d\omega_pd\omega_{p'} & A_p(\omega_p)A^*_p(\omega_{p'})e^{-i(\mathbf{q}\cdot\mathbf{r}_\alpha-\mathbf{q}'\cdot\mathbf{r}_\beta)} \\ & \notag\times \langle\hat{\sigma}(\mathbf{q},\omega'-\omega_p)\rangle \langle\hat{\sigma}(-\mathbf{q}',\omega_{p'}-\omega')\rangle
\end{align}
\begin{align}\label{eq:Sincwr}
S_{\mathrm{inc}}(\bar{\omega},\bar{\mathbf{r}},\Lambda)=K\int d\omega'\vert F_f(\omega',\bar{\omega})\vert^2 \omega'^2\sum_{\alpha}\int d\omega_pd\omega_{p'} & A_p(\omega_p)A^*_p(\omega_{p'})e^{-i(\mathbf{q}-\mathbf{q}')\cdot\mathbf{r}_\alpha}  \\ & \notag\times \langle\hat{\sigma}(-\mathbf{q}',\omega_{p'}-\omega')\hat{\sigma}(\mathbf{q},\omega'-\omega_p)\rangle
\end{align}
\end{widetext}
where $\mathbf{q}^{(\prime)}\equiv\frac{\omega'}{c}\hat{\bar{\mathbf{r}}}-\mathbf{k}_{p^{(\prime)}}$ is the momentum transfer, $F_f(\omega,\bar{\omega})$ is the frequency gating function of Appendix \ref{ap:gating} and $\Lambda$ stands for the set of parameters that define the external pulse envelopes (including $\mathbf{k}_{p^{(\prime)}}$).  We approximate
\begin{align}
K=\frac{\vert \bar{\epsilon}(\hat{\mathbf{k}}_p)\cdot\mathbf{\mu}_D\vert^2}{72\pi c^4 r'^2}
\end{align}
as a constant on the assumption that all pixels are roughly equidistant from the sample.  In approaches to
x-ray scattering that do not incorporate the detection event, the differential scattering cross section is
calculated and found to be proportional to
$r_0^2(\frac{\omega_s}{\omega_p})\vert\epsilon_p\cdot\epsilon_s\vert^2$ with $r_0$ the classical electron
radius.  Our incorporation of the detection event included a summation over polarizations of the signal field
and an averaging over initial polarizations and emission directions.  This was shown to lead to the
replacement $\epsilon_p\cdot\epsilon_s\to\bar{\epsilon}(\hat{\mathbf{k}}_p)\cdot\mu_D$ while the use of atomic
units equates $r_0^2=\frac{1}{c^4}$.  Finally, since we calculate the signal (defined as the expectation value
of the gated electric field) by explicitly incorporating the detection event (which is linear in $\omega_s$)
our result is proportional to $\omega_s^2$.  Recalling that
$\mathbf{A}(\omega)\propto\frac{1}{\omega}\mathbf{E}(\omega)$, we see that our result carries the appropriate
proportionality factors compared to the usual differential scattering cross section (\cite{schulke}).
\par
That the arguments of Eqs. (\ref{eq:Scohwr})-(\ref{eq:Sincwr}) are $\bar{\omega}$ and $\bar{\mathbf{r}}$
reflects the fact that they correspond to taking a spectrum at every pixel.  Since the final observed signal
frequency is $\bar{\omega}$, we may as well relabel it $\omega_s$ to make the interpretation more intuitive.
Aside from the expected inverse-square dependence on $\bar{r}$, the signal only depends on $\bar{\mathbf{r}}$
through $\hat{\bar{\mathbf{r}}}$, i.e., the direction vector pointing from the sample to the pixel.  Since
$\hat{\bar{\mathbf{r}}}$ is the same as the direction of propagation of scattered light, this suggests
representing the directional dependence by defining
$\frac{\omega_s}{c}\hat{\bar{\mathbf{r}}}\equiv\mathbf{k}_s$. Finally, it is important to note that, althought
these signals do not depend on time directly since we have assumed no time resolution (i.e., the pixels are
simply left open to collect incoming light), the signal does depend parametrically on the central time of the
incoming pulse through the field envelope $A_p(\omega)$ which carries a phase factor $e^{-i\omega T}$.  Here,
$T$ is the central time of the pulse envelope and the zero of time is set at the end of the state preparation
process (where the prepared state is presumed to be known).  Since $T$ therefore represents the time
separation between state preparation and arrival of the center of the probe pulse and this is a key
experimental control, we explicitly write this dependence in future expressions.

\subsection{Eigenstate Expansion of the Frequency-Resolved Signal}
In the following, we focus on the frequency-resolved signal because of its relative ease of interpretation.
While this has not yet been demonstrated in the X-ray regime, it has been shown possible to discriminate a
single wavelength component from multiwavelength scattering data in the EUV range \cite{dilanian}. From
supplementary Eqns.~(\ref{eq:Scohwr}) and (\ref{eq:Sincwr})) with $\vert
F_f(\omega',\bar{\omega})\vert^2=\delta(\omega'-\bar{\omega})$, this signal is given by the sum of a coherent
and an incoherent contribution which are related to the transition charge density $\sigma_{ab}(\mathbf{q})$
\begin{align}\label{eq:SkMixed}
S(\mathbf{k}_s,\Lambda)=K\sum_{\alpha\ne \beta} &\sum_{abcd}\rho_{ab}\rho_{cd}^*\omega^2_s  e^{i(\omega_{ba}T-\mathbf{q}_{ba}\cdot\mathbf{r}_\alpha)}e^{-i(\omega_{dc}T-\mathbf{q}_{dc}\cdot\mathbf{r}_\beta)}\mathcal{A}_p(\omega_s+\omega_{ba}) \mathcal{A}^*_p(\omega_s+\omega_{dc}) \sigma_{ba}(\mathbf{q}_{ba})\sigma^*_{dc}(\mathbf{q}_{dc})\\ \notag +K\sum_\alpha&\sum_{abc} \rho_{ac}\omega^2_se^{i(\omega_{ba}T-\mathbf{q}_{ba}\cdot\mathbf{r}_\alpha)} e^{-i(\omega_{bc}T-\mathbf{q}_{bc}\cdot\mathbf{r}_\alpha)}\mathcal{A}_{p}(\omega_s+\omega_{ba})\mathcal{A}^*_{p}(\omega_s+\omega_{bc})\sigma_{ba}(\mathbf{q}_{ba})\sigma^*_{bc}(\mathbf{q}_{bc}),
\end{align}
where $\mathcal{A}_p(\omega)$ is the spectral envelope of the scattered pulse, $\sigma_{ij}(\mathbf{q})$ are Fourier transformed matrix elements of the charge-density operator and $a$,$b$,$c$, and $d$ represent electronic states. We have also defined the momentum-transfer vector $\mathbf{q}_{ba}\equiv\mathbf{k}_s-\frac{\omega_s+\omega_{ba}}{c}\hat{\mathbf{k}}_p$.
\par
The signal is not generally related to the time-dependent, single-particle charge density but rather to its correlation function \cite{dixit, schulke}.  A compact expression for the total signal is
\begin{align}\label{eq:Stot}
&S_{T}(\mathbf{k}_s,\Lambda)=K\int d\omega'\vert F_f(\omega',\bar{\omega})\vert^2 \omega'^2\int d\omega_pd\omega_{p'}A_p(\omega_p)A^*_p(\omega_{p'})\langle\hat{\sigma}_T(-\mathbf{q}',\omega_{p'}-\omega_s)\hat{\sigma}_T(\mathbf{q},\omega_s-\omega_p)\rangle
\end{align}
where the correlation function of the total charge density operators may be expanded in terms of the single-particle densities as
\begin{align}\label{eq:cofunc}
\langle\hat{\sigma}_T(-\mathbf{q}',\omega_{s}-\omega_{p'})&\hat{\sigma}_T(\mathbf{q},\omega_p-\omega_s)\rangle =\\ \notag\sum_{\alpha}e^{-i(\mathbf{q}-\mathbf{q}')\cdot\mathbf{r}_\alpha}\langle\hat{\sigma}(-\mathbf{q}',\omega_{s}-\omega_{p'})\hat{\sigma}(\mathbf{q},\omega_p-\omega_s)\rangle+&\sum_{\alpha}\sum_{\beta\ne\alpha}e^{-i(\mathbf{q}\cdot\mathbf{r}_\alpha-\mathbf{q}'\cdot\mathbf{r}_{\beta})}\langle\hat{\sigma}(-\mathbf{q}',\omega_{s}-\omega_{p'})\rangle\langle\hat{\sigma}(\mathbf{q},\omega_p-\omega_s)\rangle
\end{align}
For a macroscopic sample initially in the ground electronic state, the standard classical theory of
diffraction gives the signal as the product of the modulus square of the single-particle momentum-space
electron density (the form factor) and a structure factor which describes the interparticle distribution
(equation (\ref{eq:SkClassicSep})) \cite{modxrayphys}.  This formula is used to invert the X-ray diffraction signal
to obtain the ground state electron density once the ``phase problem" is resolved \cite{elserPhase}. Substituting equation (\ref{eq:cofunc}) into (\ref{eq:Stot}) yields a close resemblence to the classical expression
for X-ray diffraction except that the coherent and incoherent terms now carry different form-factors.  Note that, if we restrict attention to elastic scattering, the correlation function in the incoherent term separates into a modulus square form and the two form factors are equal.

\section{Time- and Wavevector-Dependent X-Ray Scattering from a Single Cysteine Molecule}
\begin{figure}
\includegraphics[width =0.5\textwidth]{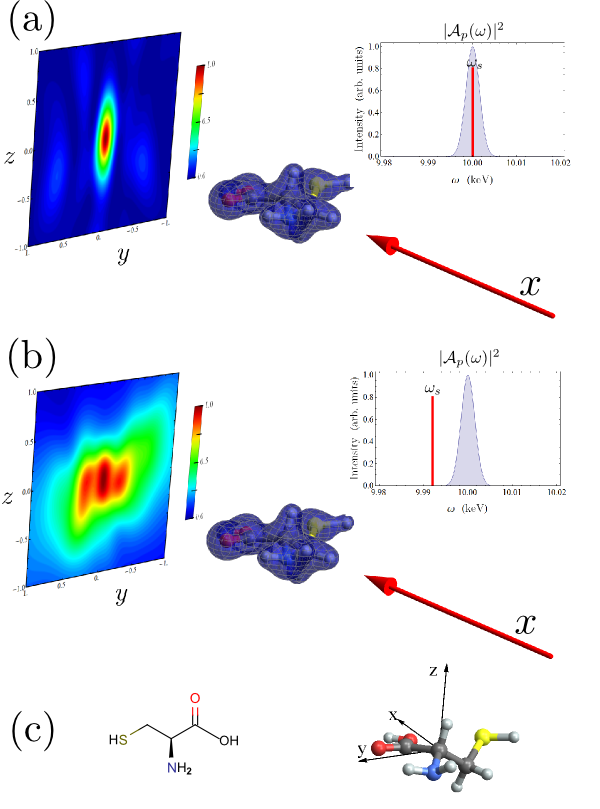}
\caption{ Off-resonant scattering of a Gaussian X-ray pulse from cysteine for different detection frequencies.  On the
right we show the pulse power spectrum in blue, with the detection frequency marked as a red line.  The pulse
propagation vector is shown as a red arrow, pointing at the molecule aligned in the lab frame, with the
scattering pattern shown in the background.  \textbf{a}: The detection frequency $\omega_s$ is set equal to
the pulse center frequency $\Omega_p$, and the scattering signal is dominated by the elastic term.
\textbf{b}: The detection frequency is set to $\Omega_p-9\,\mathrm{eV}$, and the inelastic terms are
dominant.  \textbf{c}: Chemical structure (left) and lab-frame orientation (right) of the cysteine molecule
(O is red, S is green, N is blue, C is grey, H is white.}
\label{fig:gssignal}
\end{figure}

Cysteine is a sulfur-containing amino acid which affects the secondary structure of many protiens because of the disulfide
bonds it forms.  It has been implicated in biological charge transfer in respiratory
complexes \cite{hayashi_electron_2010}.  We have previously explored various resonant X-ray spectroscopic
signals from this molecule, including stimulated X-ray Raman scattering (SXRS) and X-ray photon echo \cite{zhang:194306,biggs1.4799266}.
Below, we present calculations of off-resonant scattering from a single cysteine molecule (chemical structure and
oriention shown in Fig.~\ref{fig:gssignal}c). Details of the computational methodology can be found in section \ref{sec:methods}.

The scattering signal from the ground state (the second term in equation \ref{eq:SkMixed} with $a=c=g$, the ground state)
is depicted in Figs. \ref{fig:gssignal}\textbf{a} and \ref{fig:gssignal}\textbf{b}.  We also show the ground-state electron density, $\sigma_{gg}(\mathbf{r})$,
the pulse power spectrum, and the pulse wave vector for reference.
As these calculations are for
a single isolated molecule, we can restrict our scattering calculations to the second (incoherent) term in equation (\ref{eq:SkMixed}).
We take the scattering pulse to be a transform-limited Gaussian
\begin{equation}\label{eq:ejgauss}
 \mathcal{A}_p(\omega) = \mathcal{A}_p \sqrt{2\pi} \tau_p e^{-\tau_p^2(\omega-\Omega_p)^2/2} .
\end{equation}
The center frequency $\Omega_p$ is set to 10 keV, and we take the direction of propagation to be in the
positive $x$ direction (for molecule orientation, see Fig.~\ref{fig:gssignal}.\textbf{c}).  The pulse duration
is $\tau_p=300$ $\mathrm{as}$ which orresponds to a fwhm bandwidth of 3.65 eV.  Future progress in
pulse-generation may make such experiments realizable.
For this frequency range, the difference between $\tilde{\mathbf{q}}$ and $\mathbf{q}_{ba}$ for any two states
$a$ and $b$ is negligibly small, and is ignored in the calculations presented herein.

We take the signal detectors to be on square grid, 2 cm in length on each side, located 1 cm from
the molecule in the positive $x$ direction (i.e. we detect forward-scattered light). This corresponds to a
maximum detected scattering angle of $54.7^\circ$. We consider two different values for the detection
frequency $\omega_s$, one inside and one outside the pulse bandwidth.  When we set the detection frequency
equal to the pulse center frequency, we get the signal shown in Fig.~\ref{fig:gssignal}.\textbf{a}.  This
signal is dominated by the elastic scattering terms, where the scattering process does not change the state of
the molecule.  At this detection frequency, the elastic contribution is $4.4 \times 10^6$ larger than the
inelastic.
\par
The elastic scattering term can be eliminated by moving the detection frequency outside the pulse bandwidth.
In Fig.~\ref{fig:gssignal}.\textbf{b}, we show the scattering signal with a detection frequency
$\omega_s=\Omega_p - 9 \, \mathrm{eV}$.
With this detection frequency, we see inelastic terms from valence states $e$ whose excitation energy
satisfies the condition that  $\omega_{eg}+\omega_s$ is within the pulse bandwidth.  Therefore all states with
an energy between 4 and 12 eV will contribute.  The scattering pattern resulting from the elastic and
inelastic process are vastly different.  The former is more strongly centered around the origin, corresponding
to $q=0$, and elongated in the $z$ direction.  The inelastic term, in addition to the feature at the origin,
has two equal-intensity peaks at $(y,z)=(-0.225\,\mathrm{cm},-0.05\,\mathrm{cm})$ and
$(0.25\,\mathrm{cm},0.0\,\mathrm{cm})$, which, when converted to reciprocal space corresponds to
$(q_x,q_y,q_z) = (-0.08\,\mathrm{au}^{-1}, -0.65\,\mathrm{au}^{-1},0.0\,\mathrm{au}^{-1})$ and
$(-0.07\,\mathrm{au}^{-1}, -0.60\,\mathrm{au}^{-1},-0.13\,\mathrm{au}^{-1})$, respectively.
\par
We next turn to time-resolved scattering, in which an X-ray Raman preparation pulse $\mathcal{A}_R(\omega)$,
resonant with the sulfur K edge, arrives at $t=0$ followed by off-resonance scattering at time $t=T$.  In this
process, the Raman pulse acts twice on the same side of the loop, first promoting a sulfur 1s electron to the
valence band before the transient core hole is filled by another valence electron.  Because the Raman pulse is
broadband, these two dipole interactions leave the molecule in a superposition of valence-excited states.
This wavepacket is initially localized in the region surrounding the atom whose core is in resonance (sulfur
in this case), but becomes delocalized across the molecule in a $< 5\, \mathrm{fs}$ time scale
\cite{healion_entangled_2012,Biggs24092013}.
\par
The molecular density matrix immediately following the interaction with the first pulse is
\begin{equation}\label{eq:rho}
\hat{\rho}=i \hat{\alpha}\hat{\rho}_0 -i \hat{\rho}_{0} \hat{\alpha}^\dagger
\end{equation}
where
\begin{equation}
   \label{eq:alpha}
\hat{\alpha} = \sum_{c,e} \ket{e}
   \frac{(\boldsymbol{\epsilon}_R \cdot \boldsymbol{\mu}_{e c}) (\boldsymbol{\epsilon}_R \cdot \boldsymbol{\mu}_{c g}) }{2\pi}
   \int _{-\infty }^{\infty } d\omega \frac{\mathcal{A}_R^*\left(\omega\right)\mathcal{A}_R\left(\omega+\omega_{eg}\right)}{\omega-\omega_{ce}+i \Gamma_{c}} \bra{g}
\end{equation}
is the effective polarizability operator and $\hat{\rho}_0$ is the initial (equilibrium) density matrix.
In equation (\ref{eq:alpha}), $\boldsymbol{\epsilon}_R$ is the polarization vector for the Raman pulse and
$\boldsymbol{\mu}_{ec}$  is the transition dipole between the valence-excited state $e$ and the core-excited state $c$.
\par
For a single-molecule system prepared in this manner (and with the simplification $\mathbf{q}\to\mathbf{q}_0$), equation (\ref{eq:SkMixed}) assumes the form
\begin{equation}\label{eq:TDXRD}
S(\boldsymbol{k}_s,T) = \sum_{e,e'} i \alpha_{e,g} e^{-i \omega_{eg}T} \mathcal{A}_p(\omega_s+ \omega_{e'e})\mathcal{A}^*_p(\omega_s+ \omega_{e'g}) \sigma_{e'e}(\tilde{\mathbf{q}})\sigma^*_{e'g}(\tilde{\mathbf{q}}) + \mathrm{c.c}
\end{equation}
Note that any amplitude in the ground state after the Raman pulse has passed (terms in equation (\ref{eq:TDXRD}) where $e=g$) will contribute to a background, delay-time-independent signal, which can be filtered out.  The remaining time-dependent scattering signal is a difference signal and will have positive and negative features, unlike the ground-state scattering signals from Fig. \ref{fig:gssignal} which were only positive.
The largest contributions will come from terms where $e'$ in equation (\ref{eq:TDXRD}) is equal to either $e$ or $g$.

\begin{figure}
\includegraphics[width =\textwidth]{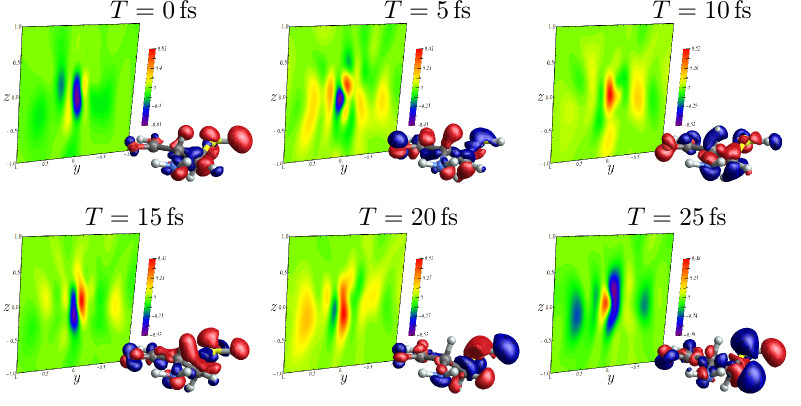}
\caption{Background: time-dependent X-ray scattering (with $\omega_s=\Omega_p$) following X-ray Raman
scattering (equation (\ref{eq:TDXRD})) for various interpulse delay times.  Foreground: real-space transition
charge densities for the Raman wavepacket (equation (\ref{eq:rsdm})). Refer to Fig. \ref{fig:gssignal}c for
the positions lab-frame orientation.}
\label{fig:TDXRD1}
\end{figure}

We take the Raman pulse center frequency at the sulfur K-edge frequency $\Omega_R = 2.473 \,\mathrm{keV}$, and
polarized along the $x$ direction. The X-ray Raman signal is highly dependent upon the choice of polarization
vector, and the nature of the underlying wavepacket is quite different for a $y$ or $z$ polarized pulses
\cite{sma_reference}. We take both the Raman and scattering pulses to be Gaussian with duration
$100\,\mathrm{as}$ (fwhm of 10.96 eV). The broad bandwidth connects the ground state with the set of valence
excited states, with energies between 5.7 eV and 9.0 eV.  Figure \ref{fig:TDXRD1} shows the time-dependent
X-ray scattering signal for interpulse delays ranging from 0 to 20 fs.
For each signal, we also show the transition density for the Raman wavepacket prior to interaction with the scattering pulse. This is defined by
\begin{equation}\label{eq:rsdm}
\mathrm{Tr} \big[\hat{\sigma}(\boldsymbol{r})\hat{\rho}\big]= \sum_{e} i \alpha_{eg} e^{-i \omega_{eg} T} \sigma_{eg}(\boldsymbol{r}) + \mathrm{c.c.}
\end{equation}

The left panel of Fig.~\ref{fig:TDXRD1} shows that the transition density is localized near the sulfur atom at $T=0\,\mathrm{fs}.$  In supplementary material we show a movie of the time-dependent scattering signal and transition density for interpulse delays up to 20 fs.
From the movie (see supplemental material \cite{MovieSupp}) and from Fig.~\ref{fig:TDXRD1} we see that there is a good deal more structure in the scattering
signal along the $y$ direction than along the $z$ direction.  This is consistent with the fact that the
electronic motion induced by the Raman pulse is mostly in the $y$ direction.  While the correspondence between
electronic motion in real space and the resulting scattering pattern is highly suggestive, it is not immediately
apparent whether the transition density can be recovered from the scattering pattern alone.  This is because the scattering pattern (equation \ref{eq:TDXRD}) is not simply the Fourier transform of the density (equation \ref{eq:rsdm}).

\begin{figure}
\includegraphics[width =\textwidth]{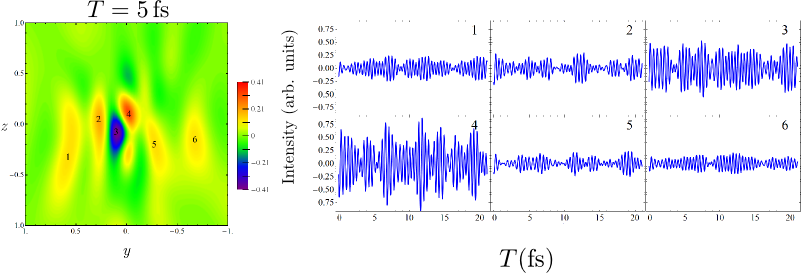}
\caption{Time-dependence of the off-resonant X-ray scattering plot (with $\omega_s=\Omega_p$).  Left: The scattering signal for $T=5\,\mathrm{fs}$, with six different features labeled.  Right: The evolution of these different features with increasing interpulse delay.}
\label{fig:TDXRD2}
\end{figure}
\begin{figure}
\includegraphics[width =\textwidth]{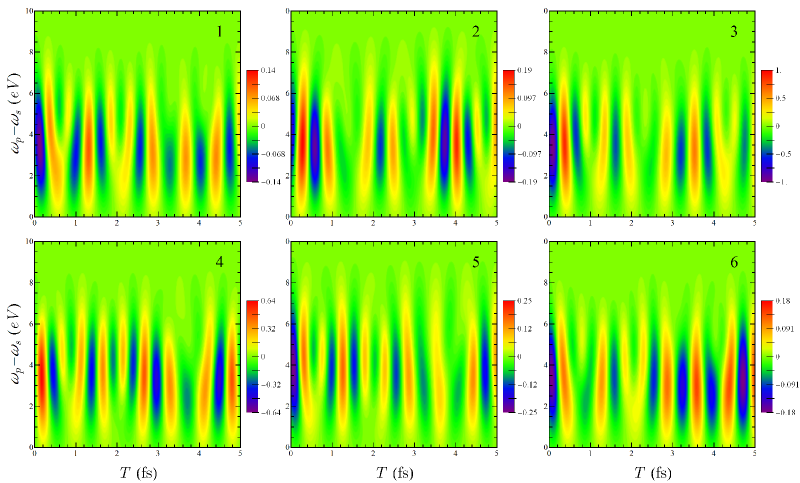}
\caption{Variation of the six features in the $T=0\,\mathrm{fs}$ scattering signal in Fig.~(5) with detection frequency $\omega_s$ and delay time $T.$}
\label{fig:8features}
\end{figure}
The scattering signal shows a complex dependence on time, reflecting interference between the many different
electronic states which make up the superposition. The signal may not simply be thought of as a snapshot of
the instantaneous time-dependent charge-density.  The time variation strongly depends on the scattering
direction, as can be seen in Fig.~\ref{fig:TDXRD2}.  Here we depict the time evolution of six points from the
$T=0\,\mathrm{fs}$ signal, corresponding to the highest and lowest peaks therein. Each trace has a beating
pattern, representing a spatially resolved interferogram.  Decay due to finite lifetime and dephasing is not included in
the time-domain signals presented here. The contribution to the signal at a given detector due to a particular
electronic coherence can be determined by Fourier transforming with respect to the delay time.  This would give
information on the transition density for the contributing excited states.  However, we do not pursue this analysis here.
\par
In Fig.~\ref{fig:8features} we show the variation of the time traces in Fig.~\ref{fig:TDXRD2} with the
detection frequency $\omega_s$.  In the previous figures, $\omega_s$ was set equal to the scattering pulse
center frequency, $\Omega_p$.  However, since purely elastic processes do not contribute to the time-dependent
signal, the signal is larger for $\omega_s<\Omega_p$.  The signal is maximized when, for a given $e$ and $e'$
from equation (\ref{eq:TDXRD}), both $\omega_s+ \omega_{e'e}$ and $\omega_s+ \omega_{e'g}$ lie within the
pulse bandwidth.

\section{Conclusions}
Coherent two-particle scattering from populations is an elastic process while coherent scattering from matter
coherences is inelastic. Incoherent scattering, in contrast, generally produces both elastic and inelastic
contributions regardless of the initial material state, as evident from equation (\ref{eq:SkMixed}).  Thus,
the coherent terms can only induce transitions between states populated by the material superposition state
while the incoherent terms can induce transitions to any electronic state.  Notably, the incoherent and
coherent signals come with different particle form factors. Thus, the total signal may not be simply factored
into the product of a particle form factor and a structure factor as in the classical theory.  This issue had
been addressed for X-ray scattering from a single hydrogen atom when it is prepared in a superposition state
\cite{dixit}. Our QED approach generalizes previous treatments \cite{schulke, dixit} to properly account for
arbitrary pulse bandshape, non-impulsive pulses, and detection details. The role of electronic coherence
requires full account of frequency, time, and wavevector gated detection as is done here.
\par
The present treatment fully incorporates inelastic scattering effects, which must be taken into account for
single-molecule scattering.  If the sample is initially in the ground state, the coherent scattering is
entirely elastic and if the sample is prepared perturbatively the coherent is dominated by elastic scattering
while the incoherent terms are affected equally by the transition charge densities ($\sigma_{eg}$).
Since light scattered from a single particle is not necessarily eleastic and can change the state of the
particle, obtaining the charge density from single-particle scattering will require distinguishing between the
Rayleigh and Raman components.  Furthermore, the total scattered intensity may not be fully described by the
ground state charge density alone; it requires more information about electronic excited states of the
particle, i.e. the transition charge densities. Our approach and simulations can provide valuable insight for
future structural studies of proteins and nano-devices.

\section{Simulation Methods}\label{sec:methods}
The details of the electronic structure calculations can be found in Ref.
\onlinecite{zhang:194306},
and are recounted briefly here.
The optimized geometry of cysteine was obtained with the Gaussian09
package\cite{G09} at the B3LYP\cite{Becke93,SDCF94}/6-311G** level of
theory. All TDDFT calculations were done at the CAM-B3LYP\cite{YTH04}/6-311++G(2d,2p) level
of theory, and with the Tamm-Dancoff approximation (TDA) \cite{HHG99b}. It was found that TDDFT with this type of long-range-corrected density functionals and diffused basis functions
can describe Rydberg states well\cite{tawada2004long, ciofini2007accurate}. In these
calculations, we
include 50 valence excited states, with energies ranging from 5.4 eV to 9.0 eV.
Core-excited states, in which a sulfur 1s electron is excited to the valence band,
were calculated
using restricted excitation windown (REW) TDDFT with a locally modified version of
NWChem
code \cite{NWChem,LKKG12}.  We also include 50 core-excited states for core excitations,
with energies ranging from 2473.5 eV
to 2495.9 eV (shifted to match experimental XANES results).

Transition density matrices between different valence excited states, which
contribute to the summation in equation (9),
are calculated using the CI coefficients from the TDDFT/TDA results, and are
therefore in an unrelaxed sense.
More accurate relaxed state-to-state transition density matrices could be calculated
using the Z-vector method \cite{Nicholas1.447489,hlwoodcock:Yamaguchi1994}, and this
research is ongoing.

\acknowledgments
The support of the Chemical Sciences, Geosciences, and Biosciences division, Office of Basic Energy Sciences, Office of Science, U.S. Department of Energy  as well as from the National Science Foundation (grants CHE-1058791 and CHE-0840513), and the National Institutes of Health (Grant GM-59230) is gratefully acknowledged.  Kochise Bennett and Yu Zhang were supported by DOE.

\appendix
\section{Time-, Frequency-, and Wavevector-Gating of Signals}\label{ap:gating}
The signal is defined as the intensity of the detected electric field
\begin{equation}
S=\int dt\int d\mathbf{r}\langle\mathbf{E}^{\dagger}(\mathbf{r},t)\mathbf{E}(\mathbf{r},t)\rangle
\end{equation}
where the detected electric field is represented as
\begin{equation}
\hat{\mathbf{E}}(\mathbf{r},t)=\frac{1}{(2\pi)^4}\int d\omega\int d\mathbf{k}e^{-i\omega t+i\mathbf{k}\cdot\mathbf{r}}\hat{\mathbf{E}}(\mathbf{k},\omega)
\end{equation}

Following the procedure outlined in Ref.~\cite{dorfman}, we add a series of gating functions to the detected electric field:
\begin{align}
&\hat{\mathbf{E}}^{(t)}(\mathbf{r},t)=F_t(t,\bar{t})\hat{\mathbf{E}}(\mathbf{r},t)\\ \notag
&\hat{\mathbf{E}}^{(t\mathbf{r})}(\mathbf{r},t)=F_{\mathbf{r}}(\mathbf{r},\bar{\mathbf{r}})\hat{\mathbf{E}}^{(t)}(\mathbf{r},t)\\ \notag
&\hat{\mathbf{E}}^{(t\mathbf{r}f)}(\mathbf{r},t)=F_f(\omega,\bar{\omega})\hat{\mathbf{E}}^{(t\mathbf{r})}(\mathbf{r},\omega)\\ \notag
&\hat{\mathbf{E}}^{(t\mathbf{r}f\mathbf{k})}(\mathbf{r},t)=F_{\mathbf{k}}(\mathbf{k},
\bar{\mathbf{k}})\hat{\mathbf{E}}^{(t\mathbf{r}f)}(\mathbf{k},\omega)
\end{align}
This gives:
\begin{align}
\hat{\mathbf{E}}^{(t\mathbf{r}f\mathbf{k})}(\mathbf{r},t;\bar{t},\bar{\omega},\bar{\mathbf{r}},\bar{\mathbf{k}})=\int d\mathbf{r}'\int dt' \hat{\mathbf{E}}(\mathbf{r}',t')F_{\mathbf{k}}(\mathbf{r}-\mathbf{r}',\mathbf{k})F_f(t-t',\bar{\omega})F_{\mathbf{r}}(\mathbf{r}',\bar{\mathbf{r}})F_t(t',\bar{t})
\end{align}
The signal is then given by equation (\ref{eq:Sgatedintensity}). We define the bare and detector spectrograms via:
\begin{equation}\label{eq:WB01}
W_B(t',\omega',\mathbf{r}',\mathbf{k}')=\int_0^{\infty}d\tau e^{-i\omega'\tau}\int d\mathbf{R} e^{i\mathbf{k}'\cdot\mathbf{R}}\langle \mathcal{T} \hat{\mathbf{E}}_{R}^{\dagger}(\mathbf{r}'+\mathbf{R}/2,t'+\tau/2) \hat{\mathbf{E}}_{L}(\mathbf{r}'-\mathbf{R}/2,t'-\tau/2)\rangle
\end{equation}
\begin{equation}\label{eq:detectspec}
W_D(t',\omega',\mathbf{r}',\mathbf{k}';\bar{t},\bar{\omega},\bar{\mathbf{r}},\bar{\mathbf{k}})=\int\frac{d\omega}{2\pi}\vert F_f(\omega,\bar{\omega})\vert^2W_t(t',\omega'-\omega,\bar{t})\int\frac{d\mathbf{k}}{(2\pi)^3}\vert F_{\mathbf{k}}(\mathbf{k},\bar{\mathbf{k}})\vert^2W_{\mathbf{r}}(\mathbf{r}',\mathbf{k}'-\mathbf{k},\bar{\mathbf{r}})
\end{equation}
Where we have defined the auxilliary functions
\begin{equation}
W_t(t',\omega,\bar{t})\equiv\int d\tau F_t^*(t'+\tau/2,\bar{t})F_t(t'-\tau/2,\bar{t})e^{i\omega\tau}
\end{equation}
and
\begin{equation}
W_{\mathbf{r}}(\mathbf{r}',\mathbf{k},\bar{\mathbf{r}})\equiv\int d\mathbf{R}F_{\mathbf{r}}^*(\mathbf{r}'+\mathbf{R}/2,\bar{\mathbf{r}})F_{\mathbf{r}}(\mathbf{r}'-\mathbf{R}/2,\bar{\mathbf{r}})e^{-i\mathbf{k}\cdot\mathbf{R}}.
\end{equation}
The signal is then given by the overlap of the two spectrograms:
\begin{equation}\label{eq:Sspectrograms}
S(\bar{t},\bar{\omega},\bar{\mathbf{k}},\bar{\mathbf{r}})=\int dt'\frac{d\omega'}{2\pi}\int d\mathbf{r}'\frac{d\mathbf{k}'}{(2\pi)^3}W_B(t',\omega',\mathbf{r}',\mathbf{k}')W_D(t',\omega',\mathbf{r}',\mathbf{k}';\bar{t},\bar{\omega},\bar{\mathbf{k}},\bar{\mathbf{r}})
\end{equation}

For brevity, the following definitions are used in the derivations:
\begin{equation}
W_D(t',\omega';\bar{t},\bar{\omega})=\int\frac{d\omega}{2\pi}\vert F_f(\omega,\bar{\omega})\vert^2W_t(t',\omega'-\omega,\bar{t})
\end{equation}
\begin{equation}
W_D(\mathbf{r}',\mathbf{k}';\bar{\mathbf{r}},\bar{\mathbf{k}})=\int\frac{d\mathbf{k}}{(2\pi)^3}\vert F_{\mathbf{k}}(\mathbf{k},\bar{\mathbf{k}})\vert^2W_{\mathbf{r}}(\mathbf{r}',\mathbf{k}'-\mathbf{k},\bar{\mathbf{r}})
\end{equation}

\section{Derivation of the Bare Spectrogram}\label{ap:barederivation}

Beginning with equation (\ref{eq:WB01}), we expand it to leading order in $H'$ (equation (1)).  This requires two interactions (one each for the ket and bra) since the signal mode is initially in a vacuum state.
\begin{align}\label{eq:WB02}
&W_B(t',\omega',\mathbf{r}',\mathbf{k}')=\sum_{\mathbf{k}_s,\mathbf{k}_{s'}}\int d\tau e^{-i\omega'\tau}\int d\mathbf{R} e^{i\mathbf{k}'\cdot\mathbf{R}}\int_{-\infty}^{t'+\tau/2}dt_1'\int_{-\infty}^{t'-\tau/2}dt_1\int d\mathbf{r}_1d\mathbf{r}_1'\langle\hat{\mathbf{E}}_{R}^{(s')\dagger}(\mathbf{r}'+\mathbf{R}/2,t'+\tau/2)\notag \\
&\times\hat{\mathbf{E}}^{(s)}_{L}(\mathbf{r}'-\mathbf{R}/2,t'-\tau/2) \hat{\mathbf{A}}^{(s')}_R(\mathbf{r}_1',t_1') \cdot\hat{\mathbf{A}}^{(p)\dagger}_R(\mathbf{r}_1',t_1')\hat{\sigma}_{T,R}(\mathbf{r}'_1,t'_1)\hat{\mathbf{A}}^{(s)\dagger}_L(\mathbf{r}_1,t_1)\cdot \hat{\mathbf{A}}^{(p)}_L(\mathbf{r}_1,t_1)\hat{\sigma}_{T,L}(\mathbf{r}_1,t_1)\rho_T(0)\rangle
\end{align}
Here, the total density matrix is the direct product of field and matter density matrices immediately following state preparation (i.e $\rho_T(0)=\rho_F(0)\otimes\rho_M(0)$).  The vector potential of the vacuum modes, $\hat{\mathbf{A}}^{(s)}(\mathbf{r},t)$, is expanded as:
\begin{equation}
\hat{\mathbf{A}}^{(s)}(\mathbf{r},t)=\sum_{\mathbf{k}_s,\nu} \sqrt{\frac{2\pi\hbar}{V\omega_s}}\epsilon^{(\nu)}(\hat{\mathbf{k}}_s)\hat{a}_{\mathbf{k}_s,\nu}e^{-i\omega_st+i\mathbf{k}_s\cdot\mathbf{r}}
\end{equation}
Here, $\hat{a}_{\mathbf{k}_s,\nu}$ is the annihilation operator for mode $s$ and polarization $\nu$, $\epsilon^{(\nu)}(\hat{\mathbf{k}}_s)$ is a unit vector in the direction of polarization, and $V$ is the field quantization volume.  The field operator is given by
\begin{equation}\label{eq:Ejs}
\hat{\mathbf{E}}^{(s)}(\mathbf{r},t)=\sum_{\mathbf{k}_s,\nu}\sqrt{\frac{2\pi\hbar\omega_s}{ V}}\epsilon^{(\nu)}(\hat{\mathbf{k}}_s)\hat{a}_{\mathbf{k}_s,\nu}e^{-i\omega_st+i\mathbf{k}_s\cdot\mathbf{r}},
\end{equation}
while the vector potential for the classical probe beam $\hat{\mathbf{A}}^{(p)}(\mathbf{r},t)$ is represented as:
\begin{equation}
\hat{\mathbf{A}}^{(p)}(\mathbf{r},t)=\sum_\nu P_\nu\epsilon^{(\nu)}(\hat{\mathbf{k}}_p)\int\frac{d\omega_p}{2\pi}A_p(\omega_p)e^{-i\omega_pt+i\mathbf{k}_p\cdot\mathbf{r}}
\end{equation}
where $P_\nu$ is the fraction of the probe pulse in polarization state $\nu$ and $\epsilon^{(\nu)}(\hat{\mathbf{k}}_p)$ is a unit vector in the direction of direction of polarization $\nu$.  Henceforth, we will use the shorthand $\sum_\nu P_\nu\epsilon^{(\nu)}(\hat{\mathbf{k}}_p)=\bar{\epsilon}(\hat{\mathbf{k}}_p)$ and assume a narrow beam so that $\hat{\mathbf{k}}_p=\hat{\mathbf{k}}_{p'}$.  Note that, by starting the $t'_1$ and $t_1$ integrations at $-\infty$, we assume that the scattered pulse is well separated from the state preparation process.  Inserting these definitions into equation (\ref{eq:WB02}), separating matter and field correlation functions (and evaluating the latter with the conditions described above), we obtain:
\begin{align}\label{eq:WB03}
&W_B(t',\omega',\mathbf{r}',\mathbf{k}')=\frac{1}{4V^2}\sum_{\mathbf{k}_s,\mathbf{k}_{s'}}\int d\tau e^{-i\omega'\tau}\int d\mathbf{R} e^{i\mathbf{k}'\cdot\mathbf{R}}\int_{-\infty}^{t'+\tau/2}dt_1'\int_{-\infty}^{t'-\tau/2}dt_1e^{i\omega_{s'}(t'+\tau/2-t_1)-i\omega_{s}(t'-\tau/2-t_1')}\notag\\
&\times \int d\omega_pd\omega_{p'}A_p(\omega_p)A_p^*(\omega_{p'})e^{-i\omega_pt_1}e^{i\omega_{p'}t_1'}\int d\mathbf{r}_1d\mathbf{r}_1'\sum_{\alpha,\beta}^{N}\sum_{\lambda,\lambda'}\left(\mathbf{\epsilon}^{(\lambda)}(\hat{\mathbf{k}}_s)\cdot\bar{\epsilon}(\hat{\mathbf{k}}_p)\right)\left(\mathbf{\epsilon}^{(\lambda')}(\hat{\mathbf{k}}_{s'})\cdot\bar{\epsilon}(\hat{\mathbf{k}}_{p})\right)\notag \\
&\times \left(\mathbf{\epsilon}^{(\lambda)}(\hat{\mathbf{k}}_s)\cdot\mathbf{\mu}_D\right) \left(\mathbf{\epsilon}^{(\lambda')}(\hat{\mathbf{k}}_{s'})\cdot\mathbf{\mu}_D\right)e^{i\mathbf{k}_s\cdot(\mathbf{r}'-\mathbf{R}/2)}e^{-i\mathbf{k}_{s'}\cdot(\mathbf{r}'+\mathbf{R}/2)}e^{-i(\mathbf{k}_s-\mathbf{k}_p)\cdot\mathbf{r}_1}e^{i(\mathbf{k}_{s'}-\mathbf{k}_{p'})\cdot\mathbf{r}_1'} \notag \\
&\times \langle \hat{\sigma}^{\beta\dagger}_R(\mathbf{r}_1',t_1')\hat{\sigma}_L^{\alpha}(\mathbf{r}_1,t_1)\rho_M(0)\rangle
\end{align}
where we have taken a dipolar-interaction model for the detection event with $\mathbf{\mu}_D$ the dipole moment of the detector.  We have also defined $\hat{\sigma}^\alpha(\mathbf{r})\equiv\hat{\sigma}(\mathbf{r}-\mathbf{r}_\alpha)$  so that $\hat{\sigma}_T(\mathbf{r})=\sum_\alpha\hat{\sigma}^\alpha(\mathbf{r})$.

\subsection{Coherent Terms}
We first examine the $\alpha\ne\beta$ terms in the above.  Assuming that the particles are uncorrelated, we have $\rho_M(0)=\rho_\alpha(0)\otimes\rho_\beta(0)$.  The correlation function therefore splits and we can separately collect factors associated with each particle.  That is, we define:

\begin{align}\label{eq:Ptd1}
\Pi^{(\alpha)}(\mathbf{r},&t)=\frac{1}{2V}\sum_{\mathbf{k}_s,\lambda}\int_{-\infty}^{t}dt_1e^{-i\omega_s(t-t_1)}\int\frac{d\omega_p}{2\pi}A_p(\omega_p) e^{-i\omega_pt_1}  e^{i\mathbf{k}_s\cdot\mathbf{r}} \notag \\
&\times \int d\mathbf{r}_1\left(\mathbf{\epsilon}^{(\lambda)}(\hat{\mathbf{k}}_s)\cdot\bar{\epsilon}(\hat{\mathbf{k}}_p)\right)\left(\mathbf{\epsilon}^{(\lambda)}(\hat{\mathbf{k}}_s)\cdot\mathbf{\mu}_D\right)e^{-i(\mathbf{k}_s-\mathbf{k}_p)\cdot\mathbf{r}_1}\langle\hat{\sigma}^{\alpha}(\mathbf{r}_1,t_1)\rangle_\alpha
\end{align}
Where $\langle\ldots\rangle_\alpha=Tr[\ldots\rho_\alpha(0)]$ is the trace over the product of the argument and the density matrix immediately after the state preparation process ($\rho_\alpha(0)$).  The coherent spectrogram is then given by:
\begin{align}\label{eq:wbcoh1}
W_{B,coh}(t',\omega',\mathbf{r}',\mathbf{k}')=\sum_{\alpha,\beta}\int d\tau e^{-i\omega'\tau}\int d\mathbf{R}e^{i\mathbf{k}'\cdot\mathbf{R}} \Pi^{(\alpha)}(\mathbf{r}'-\mathbf{R}/2,t'-\tau/2)\Pi^{(\beta)\dagger}(\mathbf{r}'+\mathbf{R}/2,t'+\tau/2)
\end{align}
In order to carry out the integration over $dt_1$ we use the Fourier Transform
\begin{equation}
\langle\hat{\sigma}^{\alpha}(\mathbf{r}_1,t_1)\rangle_\alpha=\int\frac{d\tilde{\omega}}{2\pi}e^{i\tilde{\omega} t_1}\langle\hat{\sigma}^{\alpha}(\mathbf{r}_1,\tilde{\omega})\rangle_\alpha.
\end{equation}
We thus have
\begin{align}\label{eq:Ptd2}
\Pi^{(\alpha)}(\mathbf{r},&t)=\frac{1}{2V}\sum_{\mathbf{k}_s}\sum_{ij}\bar{\epsilon}_i(\hat{\mathbf{k}}_p)\mu_{Dj}\left(\delta_{ij}-\hat{k}_{si}\hat{k}_{sj}\right) \int d\mathbf{r}_1\int\frac{d\omega_p}{2\pi}A_p(\omega_p) \\ \notag
&\times \int\frac{d\tilde{\omega}}{2\pi}\langle\hat{\sigma}^{\alpha}(\mathbf{r}_1,\tilde{\omega})\rangle_\alpha e^{i\mathbf{k}_s\cdot\mathbf{r}}e^{-i(\mathbf{k}_s-\mathbf{k}_p)\cdot\mathbf{r}_1}e^{-i\omega_s t}\int_{-\infty}^tdt_1e^{i(\tilde{\omega}+\omega_s-\omega_p)t_1}
\end{align}
where we have also expanded the dot products of the polarizations and used the identity
\begin{equation}
\sum_{\lambda}\mathbf{\epsilon}_i^{(\lambda)}(\hat{\mathbf{k}}_s)\mathbf{\epsilon}_j^{(\lambda)}(\hat{\mathbf{k}}_s)=\delta_{ij}-\hat{k}_{si}\hat{k}_{sj}.
\end{equation}
We are now free to carry out the time integration:
\begin{equation}
\int_{-\infty}^te^{i(\tilde{\omega}+\omega_s-\omega_p)t_1}=\frac{(-i)e^{i(\tilde{\omega}+\omega_s-\omega_p)t}}{\omega+\omega_s-\omega_p-i\eta}
\end{equation}
where $\eta$ is a positive infinitesimal.  We change the summation over $\mathbf{k}_s$ to an integration via
\begin{equation}
\frac{1}{V}\sum_{\mathbf{k}_s}\to\frac{1}{(2\pi)^3}\int d\mathbf{k_s}=\int\frac{\omega^2_sd\omega_s}{(2\pi c)^3}d\Omega_s
\end{equation}
and make use of the relation \cite{salam}
\begin{equation}
\int d\Omega_s\left(\delta_{ij}-\hat{k}_{si}\hat{k}_{sj}\right)e^{\pm i\mathbf{k}_s\cdot\mathbf{r}}=\left(-\nabla^2\delta_{ij}+\nabla_i\nabla_j\right)\frac{\sin{k_sr}}{k_s^3r}.
\end{equation}
This gives:
\begin{align}\label{eq:Ptd3}
\Pi^{(\alpha)}(\mathbf{r},&t)=\frac{-i }{2(2\pi)^3}\sum_{ij}\bar{\epsilon}_i(\hat{\mathbf{k}}_p)\mu_{Dj}\left(-\nabla^2\delta_{ij}+\nabla_i\nabla_j\right) \int d\mathbf{r}_1\int\frac{d\omega_p}{2\pi}A_p(\omega_p) \\ \notag
&\times \int\frac{d\tilde{\omega}}{2\pi}\langle\hat{\sigma}^{\alpha}(\mathbf{r}_1,\tilde{\omega})\rangle_\alpha e^{i\mathbf{k}_p\cdot\mathbf{r}_1}e^{i(\tilde{\omega}-\omega_p) t}\int d\omega_s\frac{\sin{\omega_s\tilde{r}/c}}{\omega_s(\omega_s-(\omega_p-\tilde{\omega}+i\eta))}
\frac{1}{\tilde{r}}
\end{align}
Where we have defined $\tilde{\mathbf{r}}=\mathbf{r}-\mathbf{r}_1$.  The $d\omega_s$ integral has poles at $\omega_s=0$ and $\omega_s=\omega_p-\tilde{\omega}$.  The term arising from the residue at the first pole will have a factor $\frac{1}{\omega_p-\tilde{\omega}}$.  Because the interaction between the sample and the field is off-resonant, $\omega_p$ will not be close to any material frequency.  The term arising from the residue of the pole at $\omega_s=0$ is negligible in such a process.  Thus, we may perform the $d\omega_s$ integration:
\begin{equation}
\int d\omega_s\frac{\sin{\omega_s\tilde{r}/c}}{\omega_s(\omega_s-(\omega_p-\tilde{\omega}+i\eta))}=\frac{\pi e^{i(\omega_p-\tilde{\omega})\tilde{r}/c}}{\omega_p-\tilde{\omega}}
\end{equation}
Using the identity
\begin{align}
\left(-\nabla^2\delta_{ij}+\nabla_i\nabla_j\right)e^{ikr}=\{(\delta_{ij}-3\hat{r}_i\hat{r}_j)(ikr-1)+(\delta_{ij}-\hat{r}_i\hat{r}_j)k^2r^2\}\frac{e^{ikr}}{r^2}
\end{align}
and rotationally averaging so that $\hat{r}_i\hat{r}_j=\frac{1}{3}\delta_{ij}$ results in
\begin{align}
&\Pi^{(\alpha)}(\mathbf{r},t)=\frac{-i(\bar{\epsilon}(\hat{\mathbf{k}}_p)\cdot\mathbf{\mu}_D)}{6(2\pi c)^2\tilde{r}}\int \frac{d\omega_p}{2\pi}A_p(\omega_p)\int \frac{d\tilde{\omega}}{2\pi}\int d\mathbf{r}_1 e^{i\mathbf{k}_p\cdot\mathbf{r}_1}e^{-i(\omega_p-\tilde{\omega})(t-\tilde{r}/c)}(\omega_p-\tilde{\omega})\langle\hat{\sigma}^\alpha(\mathbf{r}_1,\tilde{\omega})\rangle
\end{align}

Placing the origin within the sample and taking the detector to be far away (in comparison to the size of the sample) allows the approximation $\tilde{r}=\vert\mathbf{r}'-\mathbf{R}/2-\mathbf{r}_1\vert \simeq r'-\hat{\mathbf{r}}'\cdot(\mathbf{r}_1+\mathbf{R}/2)$.  Where we have substituted $\mathbf{r}=\mathbf{r}'-\mathbf{R}/2$ since that is the point at which we will eventually evaluate $\Pi^{(\alpha)}$.  Although we will later formally integrate over all $\mathbf{R}$, this represents different detection locations and thus should only be carried out over the area of a detector pixel.  The assumption that $\mathbf{R}$ is small compared to $r'$ (the distance to the detector) is thus justified.  Dropping the retardation due to $r'$ (since this uniformly delays the signal by some constant due to travel time) and replacing the $\tilde{r}$ in the denominator by $r'$  simplifies the expression yielding:
\begin{align}
&\Pi^{(\alpha)}(\mathbf{r}'-\mathbf{R}/2,t)=\frac{-i(\bar{\epsilon}(\hat{\mathbf{k}}_p)\cdot\mathbf{\mu}_D)}{6(2\pi c)^2\tilde{r}}\int d\omega_pA_p(\omega_p)\int \frac{d\tilde{\omega}}{2\pi}e^{-i(\omega_p-\tilde{\omega})(t-\frac{1}{c}\hat{\mathbf{r}}'\cdot\mathbf{R}/2)}(\omega_p-\tilde{\omega})\langle\hat{\sigma}(\mathbf{Q}(\tilde{\omega}),\tilde{\omega})\rangle e^{-i\mathbf{Q}(\tilde{\omega})\cdot\mathbf{r}_\alpha}
\end{align}
Where we have also carried out the $d\mathbf{r}_1$ integration via
\begin{align}
\int d\mathbf{r}e^{-i\mathbf{k}\cdot\mathbf{r}}\langle\hat{\sigma}^{\alpha}(\mathbf{r},\tilde{\omega})\rangle=\langle\hat{\sigma}(\mathbf{Q}(\tilde{\omega}),\tilde{\omega})\rangle e^{-i\mathbf{Q}(\tilde{\omega})\cdot\mathbf{r}_\alpha}
\end{align}
with $\mathbf{Q}^{(\prime)}(\tilde{\omega})\equiv\frac{1}{c}(\omega_{p^{(\prime)}}-\tilde{\omega})\hat{\mathbf{r}}'-\mathbf{k}_{p^{(\prime)}}$. We are now in a position to perform the integrations over $d\tau$ and $d\mathbf{R}$ in equation (\ref{eq:wbcoh1}):
\begin{align}
\int d\tau e^{-i(\omega'-\frac{\Omega}{2})\tau}=2\pi\delta(\omega'-\frac{\Omega}{2})
\end{align}
\begin{align}
\int d\mathbf{R}e^{-i(\mathbf{k}'-\frac{\Omega}{2c}\hat{\mathbf{r}}')\cdot\mathbf{R}}=(2\pi)^3\delta(\mathbf{k}'-\frac{\Omega}{2c}\hat{\mathbf{r}}')
\end{align}
with $\Omega\equiv\omega_p+\omega'_p-\tilde{\omega}-\tilde{\omega}'$ defined for convenience. The bare coherent spectrogram is then:
\begin{align}\label{eq:WBcohCF}
&W_{B,coh}(t',\omega',\mathbf{r}',\mathbf{k}')=\frac{\vert \bar{\epsilon}(\hat{\mathbf{k}}_p)\cdot\mathbf{\mu}_D\vert^2}{36(2\pi)^2c^4 r'^2}\sum_{\alpha}\sum_{\beta\ne\alpha}\int d\omega_pd\omega_{p'}d\tilde{\omega}d\tilde{\omega}'A_p(\omega_p)A^*_p(\omega_{p'})(\omega_p-\tilde{\omega})(\omega_{p'}-\tilde{\omega}')\\ \notag
&\langle\hat{\sigma}(\mathbf{Q}(\tilde{\omega}),\tilde{\omega})\rangle \langle\hat{\sigma}(-\mathbf{Q}'(-\tilde{\omega}'),-\tilde{\omega}')\rangle e^{-i\mathbf{Q}(\tilde{\omega})\cdot\mathbf{r}_\alpha}e^{i\mathbf{Q}'(-\tilde{\omega}')\cdot\mathbf{r}_\beta}e^{-i(\omega_p-\omega_{p'}+\tilde{\omega}'-\tilde{\omega})t'}\delta(\omega'-\frac{\Omega}{2})\delta(\mathbf{k}'-\frac{\Omega}{2c}\hat{\mathbf{r}}')
\end{align}

\subsection{Incoherent Terms}
The incoherent ($\alpha=\beta$) terms contain the correlation function:
\begin{equation}\label{eq:incofunc}
\langle \mathcal{T}\hat{\sigma}_R^{\alpha\dagger}(\mathbf{r}_1',t_1')\hat{\sigma}_L^{\alpha}(\mathbf{r}_1,t_1)\rangle
\end{equation}
Since there is only one operator on each side of the density matrix, there is no time ordering ambiguity and we may drop $\mathcal{T}$.  The Hilbert space expression is then
\begin{equation}\label{eq:incofunc2}
Tr[\hat{\sigma}^{\alpha\dagger}(\mathbf{r}_1',t_1')\hat{\sigma}^{\alpha}(\mathbf{r}_1,t_1)\rho_{\alpha}(0)]
\end{equation}
Although we may not factor this into a product of correlation functions, we may still go through the same series of simplifications as in the coherent case resulting in:
\begin{align}\label{eq:WBincCF}
&W_{B,inc}(t',\omega',\mathbf{r}',\mathbf{k}')=2\pi K\sum_{\alpha}\int d\omega_pd\omega_{p'}d\tilde{\omega}d\tilde{\omega}'A_p(\omega_p)A^*_p(\omega_{p'})(\omega_p-\tilde{\omega})(\omega_{p'}-\tilde{\omega}')\\ \notag
&\langle\hat{\sigma}(-\mathbf{Q}'(-\tilde{\omega}'),-\tilde{\omega}')\hat{\sigma}(\mathbf{Q}(\tilde{\omega}),\tilde{\omega})\rangle e^{-i(\mathbf{Q}(\tilde{\omega})-\mathbf{Q}'(-\tilde{\omega}'))\cdot\mathbf{r}_\alpha}e^{-i(\omega_p-\omega_{p'}+\tilde{\omega}'-\tilde{\omega})t'}\delta(\omega'-\frac{\Omega}{2})\delta(\mathbf{k}'-\frac{\Omega}{2c}\hat{\mathbf{r}}')
\end{align}
The total (incoherent plus coherent) bare spectrogram may also be written in a form similar to this
\begin{align}\label{eq:WBTotal}
&W_{B,T}(t',\omega',\mathbf{r}',\mathbf{k}')=2\pi K\int d\omega_pd\omega_{p'}d\tilde{\omega}d\tilde{\omega}'A_p(\omega_p)A^*_p(\omega_{p'})(\omega_p-\tilde{\omega})(\omega_{p'}-\tilde{\omega}')\\ \notag
&\langle\hat{\sigma}_T(\mathbf{Q}(\tilde{\omega}),\tilde{\omega})\hat{\sigma}_T(-\mathbf{Q}'(-\tilde{\omega}'),-\tilde{\omega}')\rangle e^{-i(\omega_p-\omega_{p'}+\tilde{\omega}'-\tilde{\omega})t'}\delta(\omega'-\frac{\Omega}{2})\delta(\mathbf{k}'-\frac{\Omega}{2c}\hat{\mathbf{r}}')
\end{align}
when given in terms of the total (many-particle) charge density.

\end{document}